\documentclass[a4paper,11pt]{article}
\usepackage{amssymb}
\usepackage{jcappub}
\usepackage[T1]{fontenc}

\pdfoutput=1 
\affiliation[a]{Department of Physics, Karakoram International University,\\Gilgit, Pakistan}
\affiliation[b]{Department of Mathematics, Karakoram International University,\\Gilgit, Pakistan}
\emailAdd{hunzaie@gmail.com}
\abstract{We study the quantum tunneling of scalars from charged accelerating and rotating
black hole with NUT parameter. For this purpose we use the charged Klein-Gordon equation.
We apply WKB approximation and the Hamilton-Jacobi method to solve charged the Klein-
Gordon equation. We find the tunneling probability of outgoing charged scalars from the
event horizon of this black hole, and hence the Hawking temperature for this black hole.}
\input{tcilatex}
\begin{document}

\title{Hawking radiation of scalars from accelerating and rotating black
holes with NUT parameter}
\author{Khush Jan$^{a}$ \ and H. Gohar$^{b,1}$ \note{Corresponding author}}
\maketitle

\flushbottom

\section{Introduction}

Black holes are very subtle and mysterious objects in this universe. In the
theory of general relativity, black holes are the solutions of the Einstein
field equations. Classically, black holes do not emit any type of radiations
and are perfect absorbers. During the start of 1970's, due to the remarkable
work of Bekenstein and Stephan Hawking \cite{1a, 1, 2} , the field of black
hole physics emerged in a progressive way. The relation between the laws of
thermodynamics and black hole thermodynamics \cite{22a} is very important in
this regard. It is impossible to define a temperature for black holes.
Because, every thing goes into the black hole and as a result of this, there
is no any output. If this is the case, the second law of thermodynamics
would be violated due to entry of matter, which has its own entropy, in to
the black hole which results in the decrease of the total entropy of the
universe, which contradicts the second law of thermodynamics. Bekenstein
first conjectured the relation between the properties of black hole and the
laws of thermodynamics. In 1972, again Bekenstein showed that black holes
posses entropy, $S_{bh}$, similar to the surface area of black hole, whose
increase overcomes the decrease of the exterior entropy such that the second
law of thermodynamics is preserved. He also related the surface gravity,
which is the gravitational acceleration experienced at the surface of the
black hole, with temperature of the body in thermal equilibrium. In addition
to this,1974, Stephan Hawking showed that quantum mechanically, black holes
actually radiate particles. He also showed that these radiations are purely
thermal. Virtual particles and anti particles are permanently created due to
vacuum fluctuations at the event horizon of the black hole or near the event
horizon. Here, we have three possibilities. Firstly, both the particles may
be pulled in to the black hole, secondly, both the particles may go out of
the black hole and the third possibility is one particle goes into the black
hole and other go away from black hole. For the third possibility, the anti
particle must go into the black hole to conserve energy. The particle that
has escaped becomes real and appears to be emitted by the black hole. The
anti particle, that was pulled into the black hole reduces the black hole
mass, charge and the angular momentum. As a result of this antiparticle, the
black hole finally shrinks.

Due to this breakthrough in the field of black hole physics, large number of
people started to work on these radiations. Some of the important works are
given in these references \cite{111, 222, 333}. The tunneling method \cite%
{333, 4, 5a, 6, 7}, first time used by Kraus and Wilczek, is very important
and easy method to model the Hawking radiation from black holes. Similarly,
Kerner and Mann \cite{8, 9, 10, 11} extended this work to large number of
black holes and they showed that tunnelling method is very powerful method,
which can be applied to variety of black holes. Following the work of Kerner
and Mann, recently, this method is applied to higher dimensional black holes 
\cite{11c, 11d, 11e}, black holes in String theory \cite{11b}, black strings 
\cite{6a, 6b, 6c, 6d}, accelerating and rotating black holes \cite{66, 77,
88}, dilaton black holes \cite{6f, 6g}, three dimensional black holes \cite%
{bb1, a11} and black holes with NUT\ parameter \cite{8, 15}. In these
papers, authors have discussed the emission of fermions and scalar particles
from black holes. In this way, large number of modifications are done by
different authors \cite{c1, c2, c3, c4, c5}. For example, the modification
of Hawking temperature beyond semiclassical approximation, the use of
correct coordinates to model the Hawking radiation and quantum tunneling
with back reaction. These Radiations are also discussed for black holes in
Yang-Mills theory and Kaluza Klein theory \cite{d1, d2} . In this paper, we
have applied tunneling method to model the Hawking radiation of scalars from
accelerating and rotating black holes with NUT parameter. In section two, we
have discussed black holes with NUT parameter. In section three, we have
studied the quantum tunneling of charged scalar particles from event horizon
of these black configurations. In section four, we have concluded the
important results of our work

\section{Charged accelerating and rotating black holes with NUT parameter}

Black holes with NUT (Newman-Unti-Tamburino) parameter \cite{16, 17, 18, 19}
are very important because they do not satisfy the first law of
thermodynamics unless the NUT charge of the space-time vanishes \cite{20}
and the presence of NUT charge causes a breakdown of the entropy/area
relationship \cite{21, 22}. These black holes are very important in ADS/CFT
correspondence \cite{22c, 22b}. The NUT parameter is related with the
gravitomagnetic monopole parameter of the central mass, or a twisting
property of the surrounding space-time. The line element for such black hole
is given by \cite{18}

\begin{eqnarray}
ds^{2} &=&\frac{-1}{\Omega ^{2}}\left\{ \frac{Q}{\rho ^{2}}\left[ dt-\left(
a\sin ^{2}\theta +4l\sin \frac{\theta }{2}\right) d\phi \right] ^{2}-\frac{%
\rho ^{2}}{Q}dr^{2}-\right.  \nonumber \\
&&\left. \frac{\rho ^{2}}{P}d\theta ^{2}-\frac{P\sin ^{2}\theta }{\rho ^{2}}%
\left[ adt-(r^{2}+\left( a+l\right) ^{2})d\phi \right] ^{2}\right\} .
\label{1}
\end{eqnarray}%
In expanded form, we can write the above line element as 
\begin{eqnarray}
ds^{2} &=&\frac{1}{\Omega ^{2}}\left\{ -\left( \frac{Q}{\rho ^{2}}-\frac{%
a^{2}P\sin ^{2}\theta }{\rho ^{2}}\right) dt^{2}+\frac{\rho ^{2}}{Q}%
dr^{2}+\right.  \nonumber \\
&&\left. \frac{\rho ^{2}}{P}d\theta ^{2}+\left( \frac{P(r^{2}+a^{2})^{2}\sin
^{2}\theta }{\rho ^{2}}-\frac{Qa^{2}\sin ^{4}\theta }{\rho ^{2}}\right)
d\phi ^{2}\right. -  \nonumber \\
&&\left. \frac{2a\sin ^{2}(P(r^{2}+a^{2})-Q)dtd\phi }{\rho ^{2}\Omega ^{2}}%
\right\} ,  \label{2}
\end{eqnarray}%
where 
\begin{eqnarray*}
\Omega &=&1-\frac{\alpha }{\omega }r\left( l+\cos \theta \right) ,\text{ }%
\rho ^{2}=r^{2}+\left( l+a\cos \theta \right) ^{2}, \\
P &=&1-a_{3}\cos \theta -a_{4}\cos ^{2}\theta , \\
Q &=&\left[ \left( \omega ^{2}k+e^{2}+g^{2}\right) \left( 1+2\frac{\alpha l}{%
\omega }r\right) -2Mr+\frac{\omega ^{2}k}{a^{2}-l^{2}}r^{2}\right] \times \\
&&\left[ (1+\frac{\alpha \left( a-l\right) }{\omega }r)(1-\frac{\alpha
\left( a+l\right) }{\omega }r)\right] , \\
a_{3} &=&2\frac{\alpha a}{\omega }M-4\frac{\alpha ^{2}al}{\omega ^{2}}\left(
\omega ^{2}k+e^{2}+g^{2}\right) ,\text{ }a_{4}=-\frac{\alpha ^{2}a^{2}}{%
\omega ^{2}}\left( \omega ^{2}k+e^{2}+g^{2}\right) \\
k &=&\frac{1+2\frac{\alpha l}{\omega }M-3\frac{\alpha ^{2}l^{2}}{\omega ^{2}}%
\left( e^{2}+g^{2}\right) }{\frac{\omega ^{2}}{a^{2}-l^{2}}+3\alpha ^{2}l^{2}%
}.
\end{eqnarray*}%
Here, the parameters $M$, $e,$ $g$ and $\alpha $ represent the mass,
electric charge, magnetic charge and acceleration, of the black hole
respectively. while $a$ shows a black hole rotation and $l$ is a NUT
parameter. Now the metric defined by Eq. (\ref{1}) can be written as 
\begin{equation}
ds^{2}=-f(r,\theta )dt^{2}+\frac{dr^{2}}{g(r,\theta )}+\Sigma (r,\theta
)d\theta ^{2}+K(r,\theta )d\phi ^{2}-2H(r,\theta )dtd\phi ,  \label{3}
\end{equation}%
where $f(r,\theta ),\text{\ }{g(r,\theta )},\text{\ }\Sigma (r,\theta ),%
\text{\ }K(r,\theta ),\text{\ }H(r,\theta )$ are defined below 
\begin{eqnarray}
f(r,\theta ) &=&\frac{1}{\Omega ^{2}}\left[ \frac{Q}{\rho ^{2}}-\frac{%
a^{2}P\sin ^{2}\theta }{\rho ^{2}}\right] ,\text{ }g(r,\theta )=\frac{%
Q\Omega ^{2}}{\rho ^{2}},\text{ }\Sigma (r,\theta )=\frac{\rho ^{2}}{P\Omega
^{2}},  \label{4} \\
K(r,\theta ) &=&\frac{P\left( r^{2}+\left( a+l\right) ^{2}\right) ^{2}\sin
^{2}\theta }{\rho ^{2}\Omega ^{2}}-\frac{Q\left( a\sin ^{2}\theta +4l\sin
^{2}\frac{\theta }{2}\right) ^{2}}{\rho ^{2}\Omega ^{2}},  \label{5} \\
H(r,\theta ) &=&\frac{Pa\left( r^{2}+\left( a+l\right) ^{2}\right) \sin
^{2}\theta }{\rho ^{2}\Omega ^{2}}-\frac{Q\left( a\sin ^{2}\theta +4l\sin
^{2}\frac{\theta }{2}\right) }{\rho ^{2}\Omega ^{2}}.  \label{6}
\end{eqnarray}%
The electromagnetic vector potential for these Black holes is given by \cite%
{19} 
\begin{eqnarray}
A &=&\frac{1}{a\left( r^{2}+\left( a\cos \theta +l\right) ^{2}\right) }\times
\nonumber \\
&&\left[ -er\left( adt-d\phi \left( \left( l+a\right) ^{2}-\left(
l^{2}+a^{2}\cos ^{2}\theta +2al\cos \theta \right) \right) \right) \right. 
\nonumber \\
&&\left. -g\left( l+a\cos \theta \right) \left( adt-d\phi \left(
r^{2}+\left( a+l\right) ^{2}\right) \right) \right] .  \label{7}
\end{eqnarray}%
The event horizon of the charged accelerating and rotating black hole can be
calculated by putting 
\begin{equation}
g\left( r,\theta \right) =\frac{\Delta \left( r\right) }{\sum \left(
r,\theta \right) }=0,  \label{8}
\end{equation}%
where $\Delta \left( r\right) =\frac{Q}{P},$ So we have the horizon radii 
\begin{eqnarray}
r_{\alpha _{1}} &=&\frac{\omega }{\alpha \left( a+l\right) },\text{ }%
r_{\alpha _{2}}=\frac{\omega }{\alpha \left( a-l\right) },  \label{9} \\
r_{\pm } &=&\frac{a^{2}+l^{2}}{\omega ^{2}k}\left\{ -\left( \left( \omega
^{2}k+e^{2}+g^{2}\right) \frac{\alpha l}{\omega }-M\right) \right. \pm 
\nonumber \\
&&\left. \sqrt{\left( \left( \omega ^{2}k+e^{2}+g^{2}\right) \frac{\alpha l}{%
\omega }-M\right) ^{2}-\frac{\omega ^{2}k}{a^{2}+l^{2}}\left( \omega
^{2}k+e^{2}+g^{2}\right) }\right\} .  \label{10}
\end{eqnarray}%
Here, $r_{\pm }$ represent the outer horizon and inner horizons
respectively. Other two horizons are the acceleration horizons. The angular
velocity at outer horizon can be defined as 
\begin{equation}
\Omega =\frac{H(r_{+},\theta )}{K(r_{+},\theta )}=\frac{a}{r_{+}^{2}+\left(
a+l\right) ^{2}}.  \label{12}
\end{equation}%
The line element, Eq. (\ref{3}), can be written also as 
\begin{eqnarray}
ds^{2} &=&\left. -F(r,\theta )dt^{2}+\frac{dr^{2}}{g(r,\theta )}+K(r,\theta
)\left( d\phi -\frac{2H(r,\theta )}{K(r,\theta )}dt\right) ^{2}\right. 
\nonumber \\
&&\left. +\Sigma (r,\theta )d\theta ^{2},\right.  \label{hj}
\end{eqnarray}%
where $F(r,\theta )$ is given by%
\begin{equation}
F(r,\theta )=f(r,\theta )+\frac{H^{2}(r,\theta )}{K(r,\theta )}.  \label{11}
\end{equation}%
Putting the values of $f(r,\theta ),\text{\ }K(r,\theta ),\text{\ }%
H(r,\theta )$ in above Eq. (\ref{11}), we get 
\begin{equation}
F(r,\theta )=\frac{QP\rho ^{2}\sin ^{2}\theta }{\Omega ^{2}\left[ \sin
^{2}\theta P\left( r^{2}+\left( a+l\right) ^{2}\right) ^{2}-Q\left( a\sin
^{2}\theta +4l\sin ^{2}\frac{\theta }{2}\right) ^{2}\right] }.  \label{13}
\end{equation}%
By using Taylor series, the functions, $F(r,\theta )$ and $g\left( r,\theta
\right) ,$ near the event horizon at $r=r_{+}$ are given as 
\begin{eqnarray*}
F(r_{+},\theta ) &=&\frac{\left( r-r_{+}\right) Q^{^{\prime }}\left(
r_{+}\right) P\rho ^{2}\sin ^{2}\theta }{\Omega ^{2}\left[ \sin ^{2}\theta
P\left( r_{+}^{2}+\left( a+l\right) ^{2}\right) ^{2}-\left( r-r_{+}\right)
Q^{^{\prime }}\left( r_{+}\right) \left( a\sin ^{2}\theta +4l\sin ^{2}\frac{%
\theta }{2}\right) ^{2}\right] } \\
g\left( r_{+},\theta \right) &=&\frac{\left( r-r_{+}\right) Q^{^{\prime
}}\left( r_{+}\right) \Omega ^{2}}{\rho ^{2}}.
\end{eqnarray*}%
Similarly, we can write the the metric, Eq. (\ref{hj}), near the event
horizon as%
\begin{eqnarray}
ds^{2} &=&\left. F(r_{+},\theta )dt^{2}+\frac{dr^{2}}{g\left( r_{+},\theta
\right) }+K(r_{+},\theta )\left( d\phi -\frac{H(r_{+},\theta )}{%
K(r_{+},\theta )}dt\right) ^{2}\right.  \nonumber \\
&&\left. +\Sigma (r_{+},\theta )d\theta ^{2}.\right.
\end{eqnarray}%
By defining $d\chi =d\phi -\frac{H(r_{+},\theta )}{K(r_{+},\theta )}dt,$ we
can write the above metric as%
\begin{equation}
ds^{2}=-F(r_{+},\theta )dt^{2}+\frac{dr^{2}}{g\left( r_{+},\theta \right) }%
+K(r_{+},\theta )d\chi ^{2}+\Sigma (r_{+},\theta )d\theta ^{2}.  \label{a1}
\end{equation}%
Here, we are following the method of Kerner and Mann, where they used the co
rotating frame, $\chi =\phi -\Omega _{H}t,$ to deal with quantum tunneling
at event horizon at $r=r_{+}$.

\section{Quantum tunneling}

To deal with quantum tunneling near event horizon from charged accelerating
and rotating black holes with NUT parameter (Eq. (\ref{a1})) for scalar
field $\Psi ,$ we use the charged Klein-Gordon equation%
\begin{equation}
\frac{1}{\sqrt{-g}}\left( \partial _{\mu }-\frac{iq}{\hbar }A_{\mu }\right)
\left( \sqrt{-g}g^{\mu \upsilon }(\partial _{\nu }-\frac{iq}{\hbar }A_{\nu
})\Psi \right) -\frac{m^{2}}{\hslash ^{2}}\Psi =0,  \label{a}
\end{equation}%
where $A_{\mu }$ is the electromagnetic potential, $q$ and $m$ are the
charge and mass of the particle and $g$ is the determinant of the metric
tensor, $g_{\mu \nu }.$ Here, $\hbar $ is Planck's constant. In order to
apply the WKB approximation, we assume an ansatz of the form

\begin{equation}
\Psi (t,r,\theta ,\chi )=e^{\left( \frac{i}{\hslash }I(t,r,\theta ,\chi
)+I_{1}(t,r,\theta ,\chi )+O(\hslash )\right) },  \label{b}
\end{equation}%
where, $I$ is the action for outing trajectory of the particle. Substituting
Eq. (\ref{b}) in Eq. (\ref{a}) and keeping terms only in the lowest order of 
$\hbar $ and dividing by exponential term and multiplying by $\hslash ^{2}$
gives 
\begin{equation}
0=-\frac{\left( \partial _{t}I-qA_{t}\right) ^{2}}{F(r_{+},\theta )}%
+g(r_{+},\theta )\left( \partial _{r}I\right) ^{2}+\frac{\left( \partial
_{\theta }I\right) ^{2}}{\Sigma (r_{+},\theta )}+K(r_{+},\theta )\left(
\partial _{\chi }I\right) ^{2}+m^{2}  \label{c}
\end{equation}%
If we look at the symmetries of the space-time, we have the killing vectors, 
$\partial _{t}$ and $\partial _{\chi },$ and we are dealing only with radial
trajectories for any $\theta =\theta _{\circ }$, So there exists a solution
for above partial differential equation Eq. (\ref{c}) 
\begin{equation}
I=-\left( E-\Omega _{H}J\right) t+W(r,\theta _{\circ })+J\chi .  \label{d}
\end{equation}%
Here, $E$ and $J$ are constants and are associated with the energy and
angular momentum of the particle. If we do not use the co rotating frame
then we must use the solution, $I^{\ast }=-Et+W(r,\theta _{\circ })+J\phi ,$
but we have used the co rotating frame and replaced $\phi =\chi +$ $\Omega
_{H}t$ in $I^{\ast }$ and get the solution, Eq. (\ref{d}). After plugging
Eq. (\ref{d}) in Eq. (\ref{c}) $,$ we get 
\begin{equation}
0=-\frac{\left( -\left( E-\Omega _{H}J\right) -qA_{t}\right) ^{2}}{%
F(r_{+},\theta _{\circ })}+g(r_{+},\theta _{\circ })\left( \partial
_{r}W\left( r,\theta _{\circ }\right) \right) ^{2}+K(r_{+},\theta _{\circ
})J^{2}+m^{2}.  \label{e}
\end{equation}%
or%
\begin{eqnarray}
W\left( r\right) &=&\pm \int \frac{dr}{\sqrt{g(r_{+},\theta _{\circ
})F(r_{+},\theta _{\circ })}}\times  \nonumber \\
&&\left. \left( \left( \left( E-\Omega _{H}J\right) +qA_{t}\right)
^{2}-F(r_{+},\theta _{\circ })\left( K(r_{+},\theta _{\circ
})J^{2}+m^{2}\right) \right) ^{\frac{1}{2}}\right.  \label{e2}
\end{eqnarray}%
we have to integrate this integral around the the event horizon at $r$ $=$ $%
r_{+}.$ We have simple pole at $r$ $=$ $r_{+},$ so we use residue theory\
for Semi circle and we get%
\begin{equation}
W_{\pm }\left( r\right) =\pm i\pi \frac{r_{+}^{2}+\left( a+l\right) ^{2}}{%
Q^{^{\prime }}\left( r_{+}\right) }\left[ \left( E-\Omega _{H}J\right)
+qA_{t}\right] ,  \label{aa3}
\end{equation}%
where%
\begin{eqnarray}
Q^{^{\prime }}\left( r_{+}\right) &=&\left( 2\left( \omega
^{2}k+e^{2}+g^{2}\right) \frac{\alpha l}{\omega }-2M+2\frac{\omega ^{2}k}{%
a^{2}-l^{2}}r_{+}\right) \times  \nonumber \\
&&\left. (1+\frac{\alpha \left( a-l\right) }{\omega }r_{+})(1-\frac{\alpha
\left( a+l\right) }{\omega }r_{+}).\right.  \label{aa7}
\end{eqnarray}%
From Eq. (\ref{aa3}), we have 
\begin{equation}
Im(W_{\pm })=\pm \pi \frac{r_{+}^{2}+\left( a+l\right) ^{2}}{Q^{^{\prime
}}\left( r_{+}\right) }\left[ \left( E-\Omega _{H}J\right) +qA_{t}\right] .
\label{aa9}
\end{equation}%
As the probabilities of crossing the horizon from inside to outside and
outside to inside is given by \cite{5, 6} 
\begin{eqnarray}
P_{out} &\varpropto &\exp \left( \frac{-2}{\hbar }Im(I)\right) =\exp \left( 
\frac{-2}{\hbar }\left( Im(W_{+})\right) \right) ,  \label{aa0} \\
P_{in} &\varpropto &\exp \left( \frac{-2}{\hbar }Im(I)\right) =\exp \left( 
\frac{-2}{\hbar }\left( Im(W_{-})\right) \right) .  \label{aaaa}
\end{eqnarray}%
Here, $+$ and $-$ indicate for out going and in going particles. From Eq. (%
\ref{aa3}), we have%
\begin{equation}
W_{+}=-W_{-}.  \label{132}
\end{equation}%
This means that the probability of a particle tunneling from inside to
outside the horizon is given by 
\begin{equation}
\Gamma \varpropto \frac{P_{out}}{P_{in}}=\exp \left( -\frac{4}{\hslash }%
Im(W_{+})\right) .  \label{hh}
\end{equation}%
or%
\begin{equation}
\Gamma =\exp \left( -\frac{4}{\hslash }\pi \frac{r_{+}^{2}+\left( a+l\right)
^{2}}{Q^{^{\prime }}\left( r_{+}\right) }\left[ \left( E-\Omega _{H}J\right)
+qA_{t}\right] \right) .  \label{j}
\end{equation}%
From Eq. (\ref{j}), we can say the tunneling probability depends upon the
charge, $q,$ the term, $E-\Omega _{H}J,$ which is the energy of the particle
and other parameters of the black hole. The term, $-\Omega _{H}J$, is due to
presence of ergosphere of the black hole. From Eq. (\ref{j}) and \cite{15},
we have the same tunneling probability for out going scalars and fermions.
By Comparing Eq. (\ref{j}) with Boltzmann factor, $\Gamma =$ $exp[-\beta
\left( E-\Omega _{H}J\right) ],$ where $\beta $ is inverse temperature, we
have Hawking temperature by choosing $\hslash =1$ 
\begin{equation}
T_{H}=\frac{Q^{^{\prime }}\left( r_{+}\right) }{4\pi \left( r_{+}^{2}+\left(
a+l\right) ^{2}\right) },  \label{yy}
\end{equation}%
or%
\begin{equation}
T_{H}=\frac{\left( \left( \omega ^{2}k+e^{2}+g^{2}\right) \frac{\alpha l}{%
\omega }-M+\frac{\omega ^{2}k}{a^{2}-l^{2}}r_{+}\right) (1+\frac{\alpha
\left( a-l\right) }{\omega }r_{+})(1-\frac{\alpha \left( a+l\right) }{\omega 
}r_{+})}{2\pi \left( r_{+}^{2}+\left( a+l\right) ^{2}\right) },  \label{gg}
\end{equation}%
which is consistent with previous literature \cite{15} and this temperature
is sufficiently general and reduces to all the special cases after plugging
the other parameters equal to zero. We have also taken care of the
contribution of the imaginary part of the action coming from temporal
component to the tunneling probability \cite{t2, t3, t4}. Due to this
temporal contribution, the problem of so called factor two \cite{t1} to
Hawking radiations is solved.

\section{Conclusion}

In this paper, we have studied the Hawking radiation of charged scalar
particles from charged accelerating and rotating black holes with NUT\
parameter. By using \ the Hamilton-Jacobi method we have solved the charged
Klein-Gordon equation. For this purpose, we have employed the WKB
approximation to charged Klein-Gordon equation to derive the tunneling
probability of outgoing scalar particles. At the end, by comparing with the
Boltzmann factor of energy for particle, we have derived the Hawking
temperature for these black black holes. These results are found to be
consistent with the previous literature. When $l=0,$ $k=1$ in Eq. (\ref{gg})
,we recover the Hawking temperature of the accelerating and rotating black
holes with electric and magnetic charges \cite{66, 77, 88} . When $\alpha =0$%
, the Hawking temperature of the non-accelerating black holes is recovered 
\cite{23} . For $l=0,$ $k=1,$ $\alpha =0$, the Hawking temperature of the
Kerr-Newman black hole \cite{11} is obtained which is further reduced to the
temperature of the RN black hole \cite{88c} (for $a=0$). In the absence of
charge, the temperature exactly reduces to the Hawking temperature of the
Schwarzschild black hole \cite{88d} .

\acknowledgments

We are thankful to Douglas Singleton for his valuable comments.


\end{document}